\documentclass[%
 aip,
 amsmath,amssymb,
 reprint,%
]{revtex4-1}

\usepackage{graphicx}
\usepackage{dcolumn}
\usepackage{bm}
\usepackage[utf8]{inputenc}
\usepackage[T1]{fontenc}
\usepackage{mathptmx}
\usepackage{xcolor}
\begin{document}

\preprint{AIP/123-QED}

\title{Predicting second virial coefficients of organic and inorganic compounds using Gaussian Process Regression \\}

\author{Miruna T. Cretu}
 \affiliation{Department of Chemistry, Imperial College London, London SW7 2AZ, United Kingdom}%
 \affiliation{ Fritz-Haber-Institut der Max-Planck-Gesellschaft, Faradayweg 4-6, 14195 Berlin, Germany}%
\author{Jes\'us P\'erez-R\'ios}%
 \email{jperezri@fhi-berlin.mpg.de}
\affiliation{ Fritz-Haber-Institut der Max-Planck-Gesellschaft, Faradayweg 4-6, 14195 Berlin, Germany}%

\date{\today}

\begin{abstract}

We show that by using intuitive and accessible molecular features it is possible to predict the temperature-dependent second virial coefficient of organic and inorganic compounds using Gaussian process regression. In particular, we built a low dimensional representation of features based on intrinsic molecular properties, topology and physical properties relevant for the characterization of molecule-molecule interactions. The featurization was used to predict second virial coefficients in the interpolative regime with a relative error $\lesssim 1\% $ and to extrapolate the prediction to temperatures outside of the training range for each compound in the dataset with a relative error of 2.14\%. Additionally, the model's predictive abilities were extended to organic molecules unseen in the training process, yielding a prediction with a relative error of 2.66\%. Therefore, apart from being robust, the present Gaussian process regression model is extensible to a variety of organic and inorganic compounds.

\end{abstract}

\maketitle

\section{Introduction}

The long-standing goal to establish a relationship between the behaviour of a gas and its microscopic properties has admirably been achieved by the virial equation of state.\cite{kamerlingh,lj} Besides offering a rigorous depiction of the pressure, $p(T,\rho)$ as a function of the temperature\cite{abs0}, $T$, and density, $\rho$, the virial equation is founded on a solid statistical mechanics framework.\cite{mcquarrie_stat} The virial equation,

\begin{equation}
\label{eq1}
\frac{p}{RT\rho}=1+ \sum_{i=2}^{\mathcal{N}}{B_{i}(T)\rho^{i-1}},
\end{equation}

\noindent
encapsulates the departure from ideality of a gas in an infinite series of temperature-dependent coefficients, $B_i$($T$), which correspond to the molecular interaction in isolated clusters of size $i$. $B_i$($T$) is the $i$-th virial coefficient and is related to the role of $i$-body interactions in a system. In Eq.~(\ref{eq1}) $R$ denotes the ideal gas constant and the series is truncated up to a certain cluster size $\mathcal{N}$.

Two-body interactions are the most relevant to the macroscopic properties of a gas\cite{2_body_interact}, hence $B_2$ values have been tabulated for many gases\cite{springer}. Since $B_2$ can be derived from intermolecular potentials, the latter can be obtained from experimental $B_2$ through a proper parametrisation of the potential function.\cite{bird,pot_from_b2} This is conducive to the calculation of fluid properties such as enthalpy of vaporisation\cite{mcquarrie_thermo} and transport coefficients.\cite{mcquarrie_thermo,relaxation,transport} The knowledge of $B_2$ also helps to estimate critical points\cite{criticalp} and optimum conditions for crystal growth, which would otherwise require extensive screening experiments.\cite{crystal}

When it comes to the determination of the second virial coefficient, computational cost and experimental obstacles often come into play. The theoretical approach to estimate $B_2$ from the interaction potential was developed ever since the 1930s\cite{jt_coef, fowler_1925} and is adapted nowadays to more complex potential functions. However, the process is computationally expensive for all but simple molecules. Furthermore, experimental procedures give accurate results for certain ranges of temperature, however they are faced with the challenge to acquire reliable compressibility data.\cite{barkan} In the case of empirical approaches, the law of corresponding states\cite{corresp} leads, in some cases, to very accurate results, whereas in other situations, the accuracy is low.

\begin{figure}[h]
\includegraphics[width=1\linewidth]{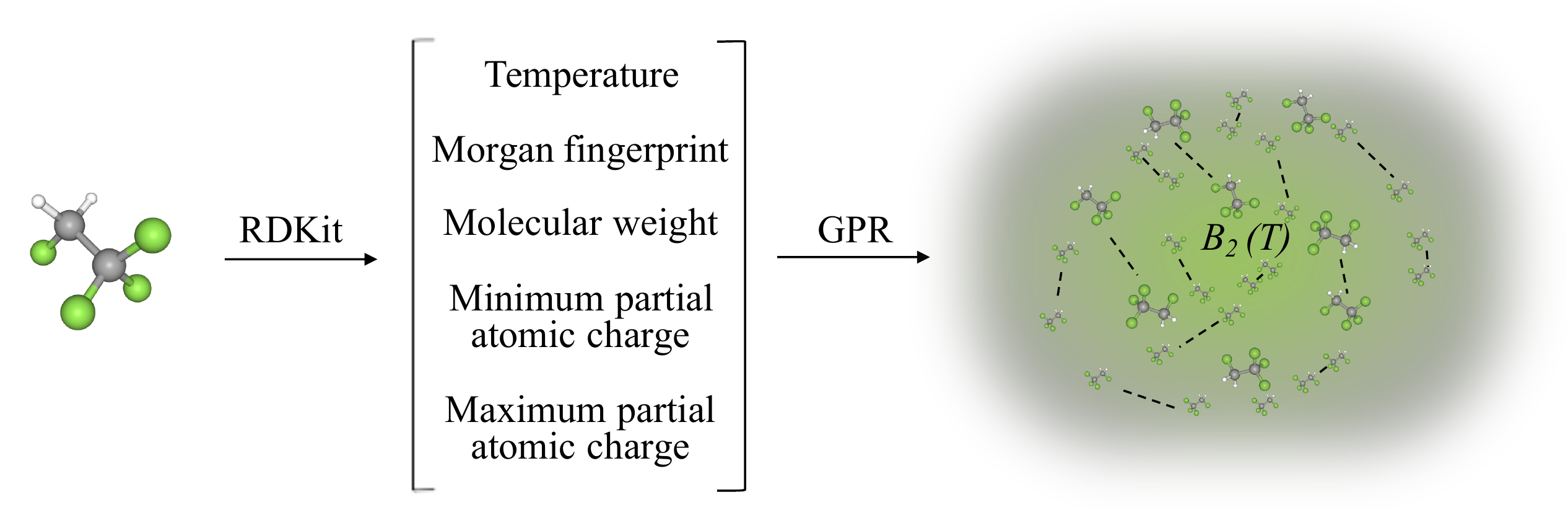}  
\caption{\label{scheme}Schematic representation of the method designed for the prediction of $B_2(T)$ using Gaussian process regression. A chosen molecule is characterized by a set of input features obtained using RDKit\cite{rdkit}. From the input data, the trained GPR model is used to predict $B_2(T)$ for the desired molecule.}
\end{figure}

To provide an alternative to the traditional methods of calculating $B_2$, we propose tackling the problem within the new paradigm of data-intensive science.\cite{datasci} The existence of a large and high quality database of temperature-dependent second virial coefficients \cite{springer} fulfills the most vital prerequisite for the application of machine learning. The choice of input features for learning is then a matter of physical and computational intuition. Among notable previous works on $B_2$ estimation is the one of Di Nicola et al.,\cite{ann} which uses thermodynamic input features and artificial neural networks (ANN) to predict $B_2$ with high accuracy. This method, however, requires the construction of a complex ANN, together with the knowledge of five thermodynamic properties, which are difficult to obtain, as discussed above. As the authors also suggest, this method should only be used when ``high accuracy is required'',\cite{ann} due to its complexity. Furthermore, the prediction of second virial coefficients has been addressed before by modelling the quantity through 2D descriptors and assuming a functional form of the dependency of $B_2$ on temperature.\cite{qspr_modelling}

\begin{figure*}
\begin{center}
 \includegraphics[width=0.7\linewidth]{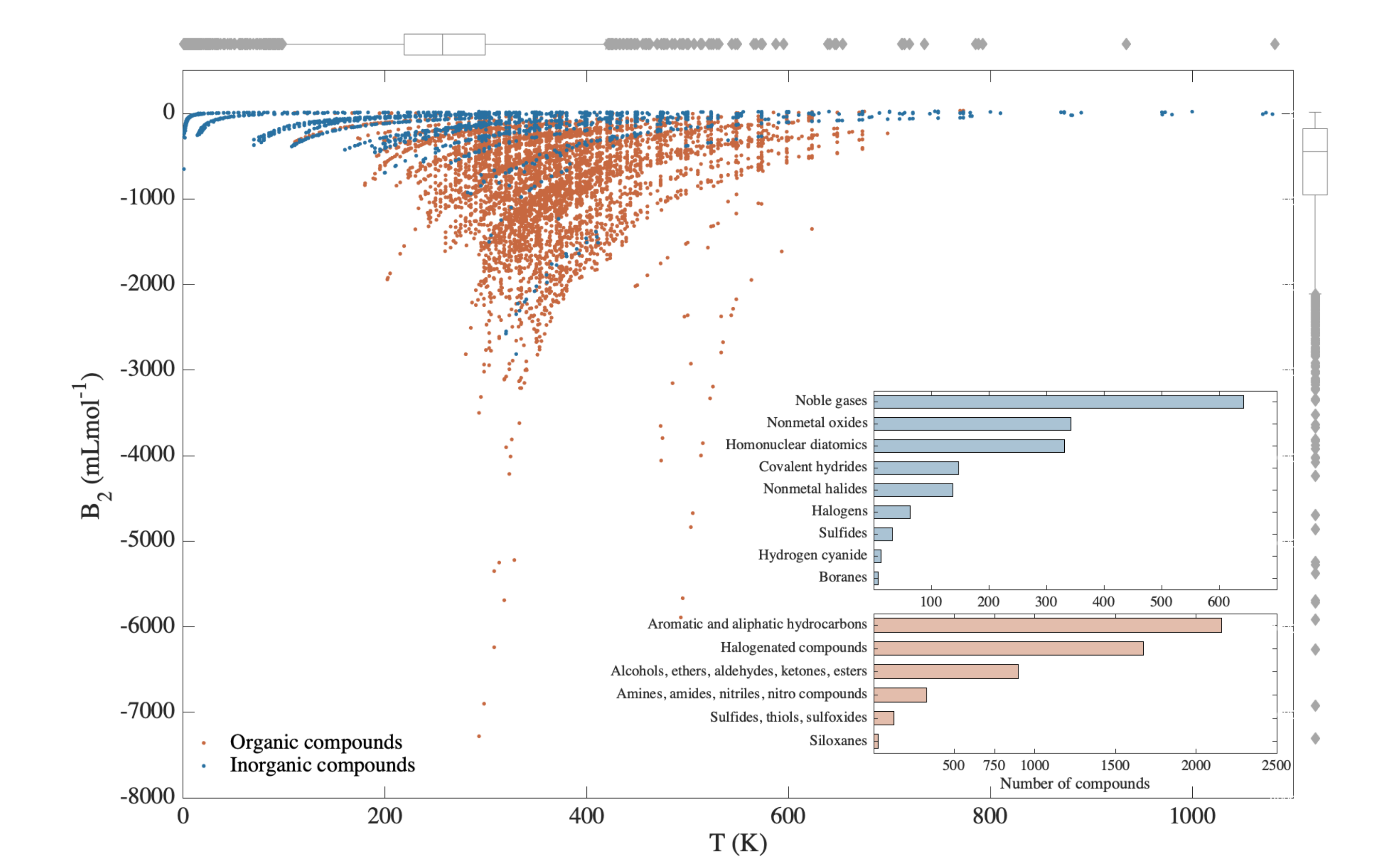} 
\caption{\label{fig:all_data} Experimental values of second virial coefficients from the filtered database as a function of temperature. The two bar charts, in the inset, show classifications of inorganic and organic compounds in the dataset, as well as the number of data points for each class of compounds. The associated box plots for temperature and second virial coefficients values are also shown, with a 1.5 maximum whisker length.}
\end{center}
\end{figure*}

In this paper, we propose the prediction of second virial coefficients of organic and inorganic compounds in a simple, universal manner. Our approach is based on Gaussian Process Regression (GPR) fed with a low dimensional input featurization scheme (see Fig.~\ref{scheme}).  

We show that the chosen features succeed to incorporate the most relevant characteristics of the second virial coefficient, in that the model succeeds to predict $B_2$ of an unseen molecule with a relative error of 2.66\%, over whole ranges of temperatures. Moreover, the power of our model is reinforced by the successful extrapolation (relative error 2.14\%) to temperatures outside of the training range for any molecule in the dataset. Our method's universality stems from its applicability to compounds belonging to a wide range of families and from the availability and accessibility of input features for any compound. The simplicity stems from the facile practice to generate input features and from the ease of applying computationally inexpensive GPR (for the number of data points considered in this work). Different featurization combinations were tested to yield the best, lowest dimensional scheme finally. All the input data were generated using RDKit,\cite{rdkit} an open-source toolkit for cheminformatics implemented in Python. While most of the features we used are basic molecular properties of compounds, the Morgan fingerprint is a representation of the connectivity of atoms in a molecule.\cite{descriptors} This mostly caters for molecular characterisation and for identifying common fragments within different molecules.

\section{The Dataset}
A comprehensive database of second virial coefficients for pure organic and inorganic substances is made available through the compilations of Dymond et al. and Gmehling et al.,\cite{data} totalling over 9300 values for a temperature range from 0.63 to 1473.15 K. It is worth emphasizing that each compound has data for the second virial coefficient in a particular range of temperatures compatible to the inherent thermo-physico-chemical properties of the compound under consideration. Subsequent to filtering, our dataset comprises 1720 data points for inorganic and 5213 data points for organic compounds, which are divided in diverse types of classes (see Fig.~\ref{fig:all_data}). While for some compounds, experimental values of $B_2(T)$ were reported for more than 200 temperatures, for other substances there existed only one data point in the set. When different $B_2$ values were registered for the same compound at the same temperature, an average of the $B_2$ values was taken. Further filtering of the data was performed by leaving out compounds with less than 3 data points and by eliminating the values which were off the temperature-dependent trend. 

The diversity of data is notable with regard to the physical and chemical properties of molecules. For instance, the inorganic compounds cover a broad spectrum of molecules and atoms starting from noble gas atoms to polyatomic molecules such as boranes. Whereas within the organic compounds, one finds ketones, which have important industrial applications,\cite{Ketones} carbonyl compounds that appear as a natural product of pollution\cite{Carbonyl} or siloxanes: an incredibly versatile class of molecules that has been proposed as a candidate for Bethe-Zel’dovich-Thompson fluids,\cite{Fluids} or that shows exciting properties as a surfactant.\cite{Hill1997}

\section{Machine Learning model}

\subsection{Gaussian Process Regression}

In the context of solving non-linear regression problems, Gaussian process regression (GPR) can be viewed as a non-parametric approach. In other words, GPR does not assume any functional form to find the fitting to a given data set. Rather, GPR employs a Gaussian distribution of functions to match the observed variables. Next, Bayesian inference, i.e., the estimation of the probability of an event given the occurrence of a previous one, allows a prior distribution of data to develop into a posterior one. In the case of GPR, the prior distribution, $p(f|\textbf{x})$ is a multivariate Gaussian distribution, usually with a zero mean function $m(\textbf{x})$ and with a covariance matrix defined by a kernel designated by the user, $K(\textbf{x},\textbf{x}')$, which stores information about the correlation between the input points.\cite{rasmussen} The distribution of functions is therefore defined as:
\begin{equation}
\label{eq2}
f(x)\sim GP(m(\textbf{x}),K(\textbf{x},\textbf{x}')).
\end{equation}
The posterior distribution, $p(f|\textbf{x},y)$, which is also normal multivariate, is obtained by conditioning the joint Gaussian prior distribution on the observations ($y$). This allows to make predictions ($y_*$) for new, unobserved data.

\subsection{Featurization methods}

The choice of input features for our model was primarily guided by chemical and physical intuition, as well as by domain knowledge. While there exists a large number of universal descriptors for machine learning which capture targeted information about systems (e.g. features related to the fitting of potential energy surfaces, such as Coulomb matrices\cite{ml_advances}), we show that good accuracy can also be obtained with the use of physico-chemical properties and molecular fingerprints solely selected using domain intuition and judgment, rather than canonical feature selection algorithms that may lead to unphysical descriptors for the problem at hand. This was proven previously in the work of Liu et al. in predicting dipole moments of diatomic molecules, explaining the good performance of intuitively chosen predictors over abstract, general purpose ones, when using small datasets.\cite{dipole_moments} To subsequently validate the chosen featurization scheme, an embedded type feature selection mechanism is implemented (see section \ref{Results}), which learns feature importance as part of the model learning process. A comparison of the performances of various combinations of predictors is also provided. 

That being said, we devised what properties would be most relevant to describing the second virial coefficient, from a physical perspective. The features used in this work belong to three categories: physical properties that can describe molecular interactions (partial atomic charges and valence electrons), topology features that characterize similarity and complexity of compounds, deduced from cheminformatics (Morgan and E-state fingerprints) and intrinsic properties of molecules (molecular weight), to account for their different sizes. All of the input features are available and easy to compute, and in our case, they were generated using RDKit.\cite{rdkit} A further explanation for the choice of physical and topology features is outlined below.

\begin{itemize}
    \item The minimum and maximum partial charges of a molecule are correlated with the molecule's dipole moment. This is supported by the recent work of Veit et. al, which implements a partial-charge model to predict dipole moments of molecules.\cite{ceriotti} The presence of a dipole moment in a molecule leads to a dipole-dipole interaction apart from the van der Waals interaction of non-polar molecules. The dipole moment of a molecule is proven to increase the attractive forces between molecules and therefore to lower $B_2$ for a given temperature.\cite{dipole_mom} This shows a direct relationship between $B_2$ and the magnitudes of the minimum and maximum partial charges. 
    
    \item Morgan fingerprints represent a well-known method for molecular characterization in terms of topology and connectivity within a molecule. In particular, a molecule is characterized by a fingerprint that contains 1024 bits, and each of these bits represents a fragment, i.e, a possible scenario of individual atoms and their environment (meaning all neighbouring atoms within a diameter of four chemical bonds) within the molecule. The ``extended connectivity'' of atoms is computed using Morgan’s extended connectivity algorithm.\cite{descriptors} Therefore, the complexity of a molecule can be assessed by counting how many bits out of 1024 are needed to describe connectivities in a molecule, as well as element types, charges and atomic masses.\cite{descriptors} Furthermore, Morgan fingerprints can be used to generate a similarity score to a reference molecule. This can be obtained through commands implemented in RDKit.\cite{rdkit} In our study, the reference molecule was chosen to be the one with the highest number of nonzero bits in the Morgan fingerprint, i.e., the most complex molecule from this point of view: 2-Ethylthiophene. In this way, a similarity score to the fingerprint of the reference molecule is attributed to each molecule in the database, as an input feature. Intuitively, this is a measure of comparison between environments, connectivities and chemical features within different molecules, which can further be relevant to comparing 2-body interactions for different molecules.
   
    \item The E-state fingerprint has also been used to characterize the molecules in the data set. This fingerprint is based on the electrotopological state indices of atoms within a molecule.\cite{descriptors} These encode information related to the valence state, electronegativity of atoms and the molecule's topology. In particular, we translate the information for each molecule into a numerical descriptor through the ratio between the total summation of E-state indices for all atoms and the summation of the number of times each possible atom type appears in the molecule.
\end{itemize}

\subsection{Model performance evaluation}

GPR, as a general fitting approach, needs a method to characterize its performance. In other words, an error estimation is needed for the proper evaluation of GPR models and the posterior identification of outliers of the model. 

One of the most common error estimators is the mean absolute error (MAE), defined as 

\begin{equation}
\label{eq5}
\text{MAE}=\frac{1}{N}\sum_{i=1}^{N}\lvert y_i-y^*_i \rvert,
\end{equation}

\noindent
where $N$ is the total number of values in the data set, $y_i$ are the true values of second virial coefficients, and $y^*_i$ are the predictions. 

In GPR, the predictions are being made after examining correlations between input features and observations in the training set. This is done without prior knowledge of the test set and, implicitly, no weighting on it, making the root mean squared error (RMSE), which is defined as follows:

\begin{equation}
\label{eq3}
\text{RMSE}=\sqrt{\frac{1}{N}\sum_{i=1}^{N}(y_i-y^*_i)^2},
\end{equation}

\noindent
a practical evaluation tool. The RMSE of predictions on the test data will be used along this work. However, when predicting physical or chemical quantities, it may be better to have a dimensionless error estimator. The normalized error ($r_E$), given as

\begin{equation}
\label{eq4}
r_E=\frac{\text{RMSE}}{y_{\text{max}}-y_{\text{min}}},
\end{equation}

\noindent
does not have units since it is defined as the ratio between the RMSE and the extension of the data. Therefore, the normalised error is an important error estimator regarding GPR, and it will be used throughout this work.

\section{Results}\label{Results}

Second virial coefficients are learned at a given temperature through a GPR model from molecular and cheminformatics-based properties of compounds. A filtered dataset of 6933 experimentally obtained second virial coefficients for different ranges of temperatures was used to train and test the GPR model. All data was randomly divided into train and test sets to find the best featurization scheme for our predictions and to investigate the performance of the chosen model, together with its covariance function and parameters. For model selection and to avoid over-fitting, 5-fold cross validation was implemented. The model's extrapolation capabilities were evaluated by testing on temperatures outside of the training range for each compound.  Finally, we challenged the limits of our model by predicting $B_2(T)$ values for molecules unseen in the training set. 

\subsection{Featurization performance analysis}

\begin{figure}
\begin{center}
 \includegraphics[width=0.8\linewidth]{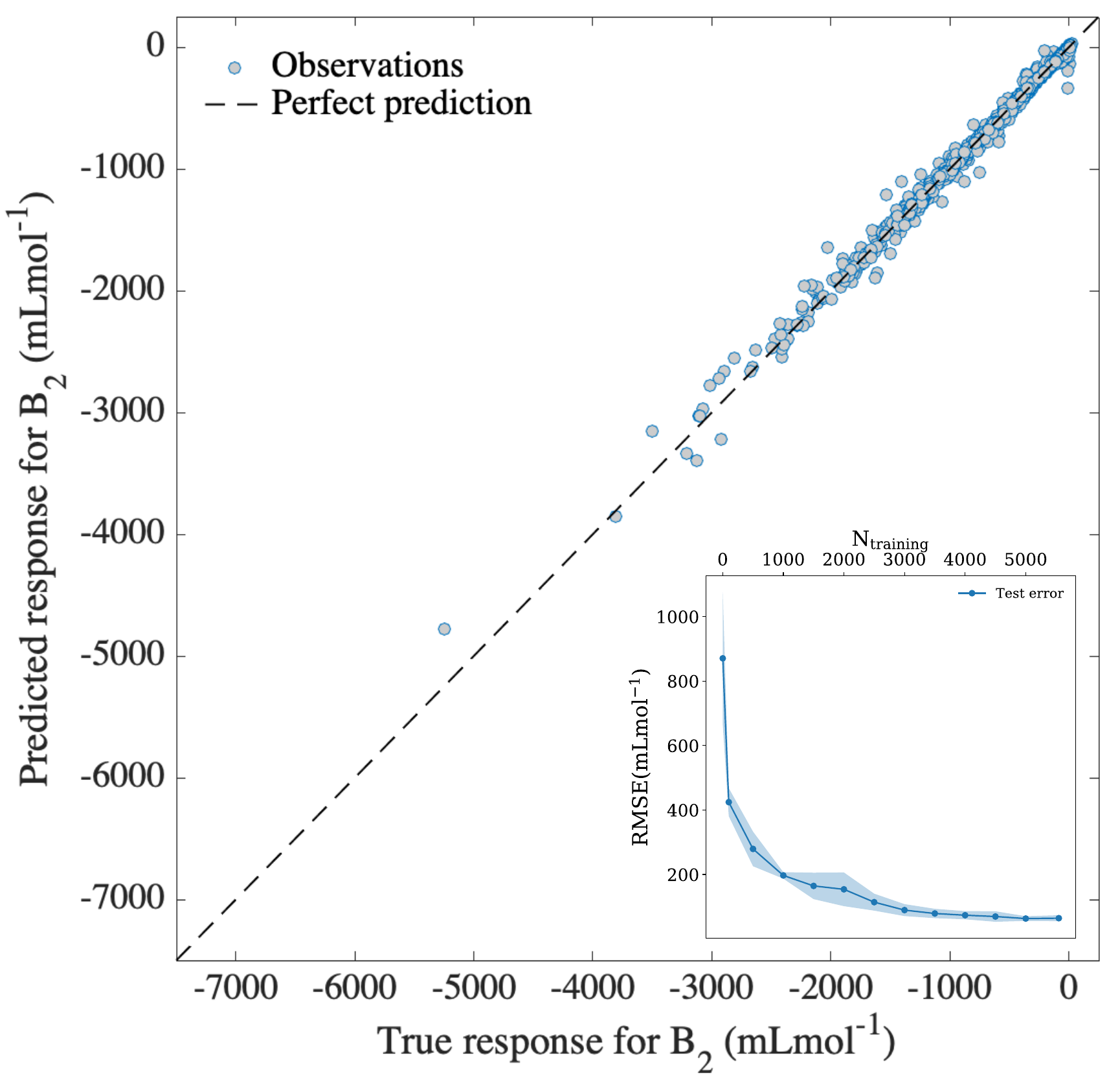}
\caption{\label{fig:best_model} GPR 5-fold cross-validated predictions of temperature-dependent second virial coefficients, $B_2(T)$, on "out-of-sample" randomly chosen test data, representing 20\% of the entire dataset. A 5 dimensional representation of input features is used (temperature, molecular weight, minimum and maximum partial atomic charges and similarity of Morgan fingerprint to that of a reference molecule). The inset shows the corresponding learning curve for this model, in which 1386 "out-of-sample" randomly selected test points were used. The shaded area stands for the error bars after 5 different iterations.}
\end{center}
\end{figure}

To assess the performances of the proposed features and to decide on the best featurization scheme, a GPR model based on a rational quadratic kernel (see Appendix) was implemented. The results of using temperature, molecular weight, minimum and maximum partial atomic charges, and similarity of the Morgan fingerprint to that of a reference molecule, as input features, are shown in Fig.~\ref{fig:best_model}. The training was done on 5547 randomly selected data points, using 5-fold cross-validation, and 1386 predictions were made on the remaining "out-of-sample" points. It is easily noticed from the figure that most of the predicted values for the second virial coefficient agree with the true experimental values, translating into an excellent performance and predictive capability of the model at hand. This astonishing performance is characterized by an RMSE of 59.99 mLmol$^{-1}$ and a normalized error of 0.82~$\%$ as it is shown in Table~\ref{table:predictors}. 

\begin{table*}
\begin{center}
\caption{\label{table:predictors}Predictors ranking by the Test RMSE score of the 5-fold cross-validated GPR model. The symbols used in the table are assigned as follows: $T$ is the temperature, $MW$ stands for the molecular weight, $\delta_{\text{min}}$, $\delta_{\text{max}}$ are minimum and maximum partial atomic charges, respectively, MF\textsubscript{nonzeros} is the number of nonzero bits in the Morgan fingerprint, MF\textsubscript{similarity} is the similarity of the compound's fingerprint to that of the reference compound, E-state encodes information on the E-state fingerprint and $VE$ is the number of valence electrons. The results are obtained using GPR trained on 5547 randomly selected training points with 5-fold CV, being tested on 1386 "out-of-sample" points.}
\begin{ruledtabular}
\begin{tabular}{c c c c c}
Dimension&Features&Test RMSE (mLmol$^{-1}$)&Test MAE(mLmol$^{-1}$)&Test $r_E$ ($\%$)\\
\hline
4D & ($T, MW, \delta_{\text{min}}, \delta_{\text{max}}$) & 92.03 & 27.37 & 1.26\\
5D & ($T, MW, \delta_{\text{min}} \delta_{\text{max}}$, MF\textsubscript{similarity}) & 59.99 & 23.34 & 0.82\\
5D & ($T, MW, \delta_{\text{min}}, \delta_{\text{max}}$, MF\textsubscript{nonzeros}) & 60.01 & 22.94 & 0.82\\
5D & ($T, MW, \delta_{min}, \delta_{\text{max}}$, E-state) & 61.87 & 23.85 & 0.85\\
5D & ($T, MW, VE$, MF\textsubscript{nonzeros}, MF\textsubscript{similarity}) & 156.11 & 53.57 & 2.13\\
6D & ($T, MW, VE, \delta_{\text{min}}, \delta_{\text{max}}$, MF\textsubscript{nonzeros}) & 56.88 & 21.87 & 0.78\\
6D & ($T, MW, VE, \delta_{\text{min}}, \delta_{\text{max}}$, MF\textsubscript{similarity}) & 66.34 & 23.47 & 0.91\\
\end{tabular}
\end{ruledtabular}
\end{center}
\end{table*}

To further analyze the performance of our GPR model we have calculated its learning curve, i.e., the model performance as a function of the number of points in the training set while keeping the number of data points in the test set constant, which is shown in the inset of Fig.~\ref{fig:best_model}. As a result, it is observed that the model's learning capabilities are converged around 4000 data points of the training set. Therefore, the performance of our model cannot benefit from having a larger dataset.

\begin{figure}
\begin{center}
 \includegraphics[width=0.75\linewidth]{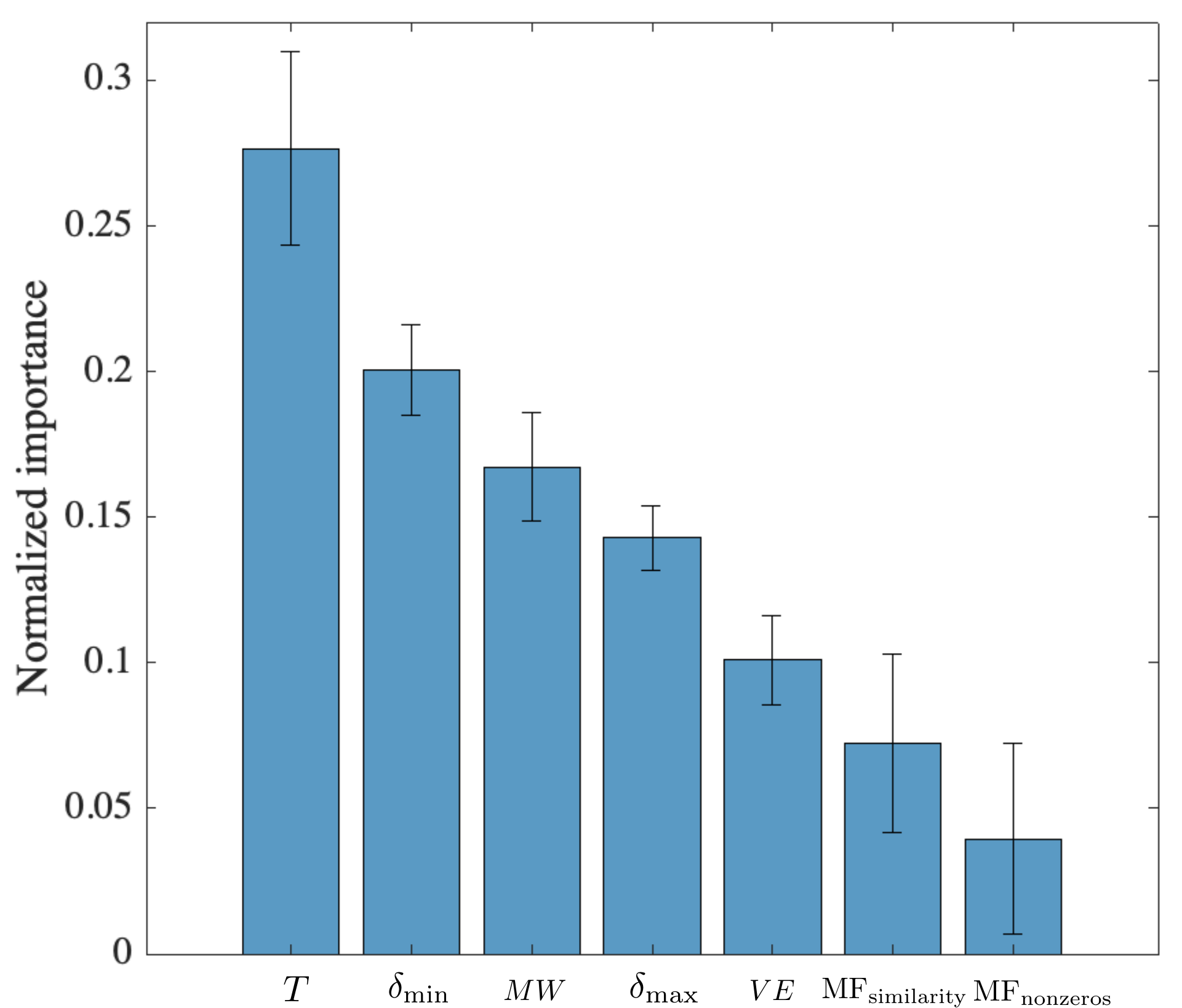} 
\caption{\label{fig:predictors_ranking} Ranking of predictors based on the characteristic length scale of each predictor, obtained with an ARD rational quadratic kernel (see Appendix). The symbols used in the figure were defined in the caption of Table \ref{table:predictors}. The errors associated with each weight are the result of performing 5 iterations.}
\end{center}
\end{figure}

The combination and the number of input features for our model were selected after the implementation of different featurization schemes and the comparison of their performances. The results of this procedure are shown in Table \ref{table:predictors}. Here, it is noticed that when the number of valence electrons is used as a predictor instead of the partial atomic charges, a much poorer performance is obtained, at the same dimensionality (5D). This is suggestive of the importance of minimum and maximum partial atomic charges as predictors in our model, presumably succeeding to account for the strength of interactions between molecules, more than just for their internal electronic structure. In addition, we notice that although the E-state fingerprint contains additional information concerning the valence state of atoms, it does not show an improved performance to that of the Morgan fingerprint in a 5-dimensional representation. Indeed, this correlates with our previous statement about the major role of partial charges in comparison with the number of valence electrons regarding molecular interactions.

To get a measure of the importance of individual predictors relative to each other, the automatic relevance determination (ARD) rational quadratic kernel function (see Appendix) was used in GPR. ARD allows the assignment of separate length scales for each predictor, instead of the same one for all of them. If an input's length scale is large, the distance one needs to move in the input space so that the function values become uncorrelated is also large, so that the covariance will become almost independent of that input. This is known as an embedded method for feature selection, as the selection is done during model training. The predictor data is standardized to allow for consistency. In this way, a weight is assigned to each input feature, as shown in Fig.~\ref{fig:predictors_ranking}. The ranking is consistent with our previous evaluation of the featurization schemes' performances (see Table \ref{table:predictors}): temperature is the most important, followed by partial atomic charges and/or molecular weight. Morgan fingerprint similarity is expected to perform better than the number of nonzero bits in the fingerprint. It is worth noticing that partial charges are better ranked than the number of valence electrons. This confirms our intuition on the relevance of partial charges over number of valence electrons, which is also supported by the results shown in Table~\ref{table:predictors}.

\subsection{Extrapolation to marginal temperatures}

\begin{figure}
\begin{center}
 \includegraphics[width=0.8\linewidth]{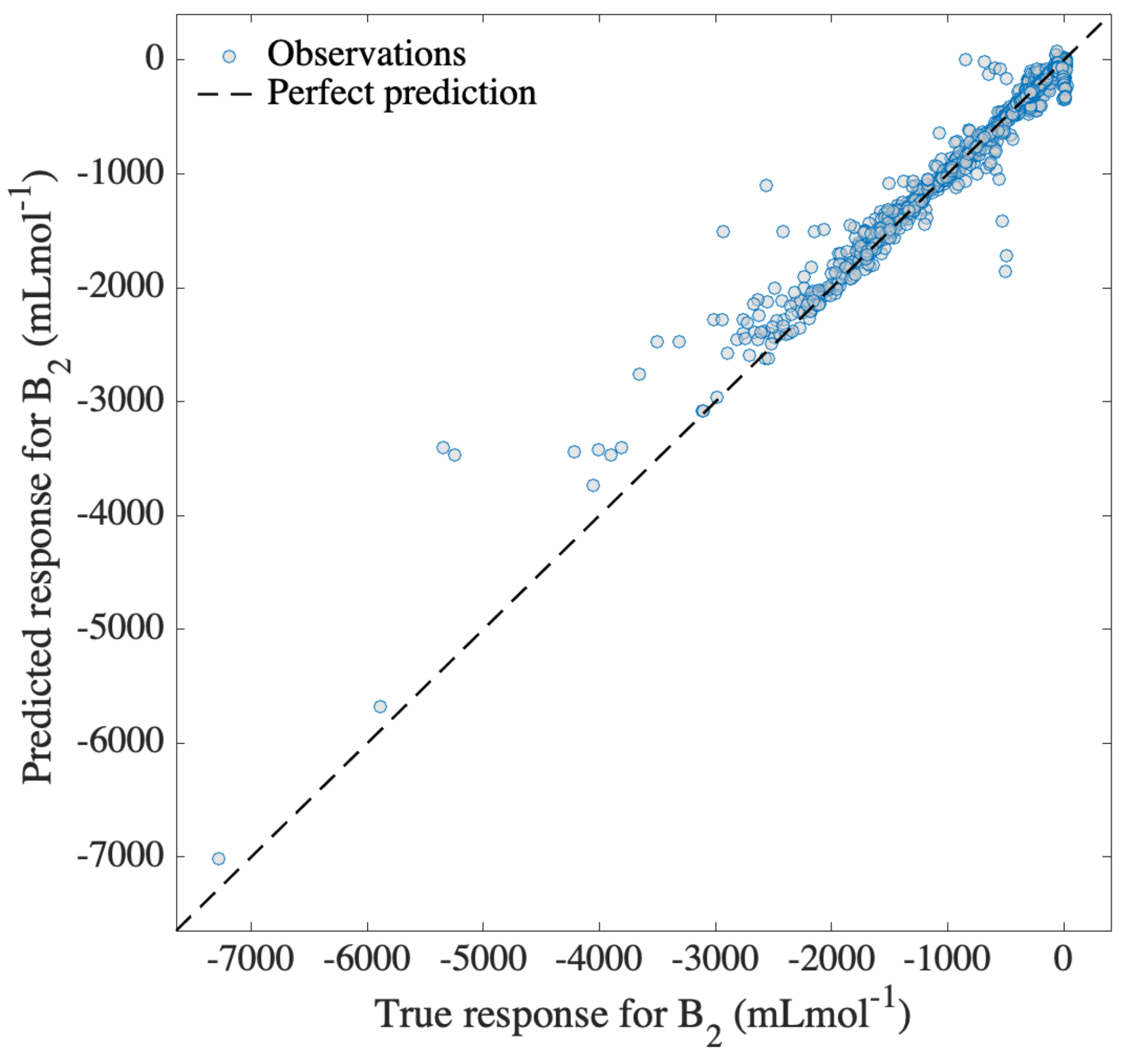} 
\caption{\label{fig:marginal_temp} GPR 5-fold cross-validated predictions of $B_2(T)$ for marginal temperatures of each compound (representing 20\% of the entire data set). A 5 dimensional representation of input features is used (temperature, molecular weight,  minimum and maximum partial atomic charges and similarity of Morgan fingerprint to that of a reference molecule).}
\end{center}
\end{figure}

To evaluate our model's extrapolation capability, the data was divided as follows: for each molecule, data points corresponding to marginal temperatures (meaning the lowest 10\% and the highest 10\% temperatures) were used for testing, whereas the rest were used for training. This selection naturally yielded a training set comprising 80\% of the total data, on which 5-fold cross-validation was applied. A GPR model based on a rational quadratic kernel was implemented, using the best featurization scheme obtained in the previous subsection (temperature, molecular weight,  minimum and maximum partial atomic charges and similarity of Morgan fingerprint to that of a reference molecule). This allowed a prediction of second virial coefficients characterized by an RMSE of 157.15 mLmol$^{-1}$ and by a relative error of 2.14\%, which is portrayed in Fig. \ref{fig:marginal_temp}. By this means, it can be noticed that the model achieves successful extrapolation for low, as well as for high temperatures corresponding to compounds in our dataset, with a few exceptions. Among these, the most distinguishable outliers are generated by ethanenitrile, toluene, methyl ethanoate and ethanol. 

While Gaussian process regression serves well as an approach to smooth interpolation,\cite{rasmussen} it usually performs worse in extrapolation tasks. Our model illustrates well the former statement, exhibiting excellent interpolation in the previous subsection. In addition to this, the representation of input features, as well as the chosen kernel function prove to lead to reasonable extrapolation capability, with only a few molecules experiencing the limitations of the model. The rational quadratic kernel (see Appendix) differs from the squared exponential one (a popular choice of kernel function for GPR) in that it contains an additional parameter ($\alpha$) which determines the relative weighting between large and small-scale variations.\cite{rasmussen} This eventually leads to a better generalization of the long term trend, compared to a squared exponential kernel, and therefore better extrapolation. The fact that the same model achieves both interpolation and extrapolation translates into considerate choice of both input features and kernel function.

\subsection{Applicability and transferability}

\begin{figure}
\begin{center}
 \includegraphics[width=0.8\linewidth]{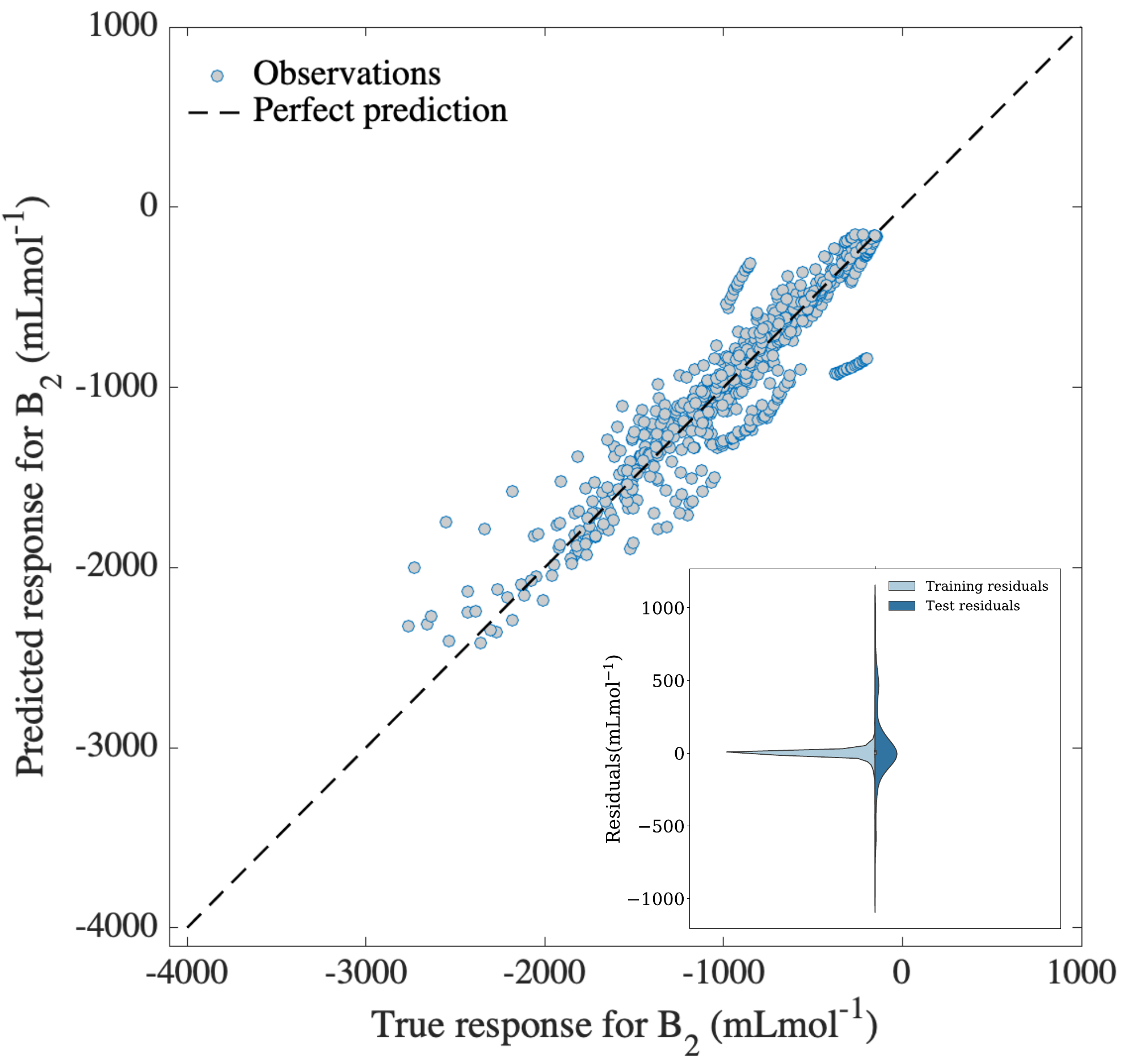} 
\caption{\label{fig:transferability} GPR 5-fold cross-validated predictions of $B_2(T)$ for organic molecules absent in the training set. A 5 dimensional representation of input features is used (temperature, molecular weight,  minimum and maximum partial atomic charges and similarity of Morgan fingerprint to that of a reference molecule). The inset shows a violin plot of residuals for the training and test sets.}
\end{center}
\end{figure}

\begin{figure}
\begin{center}
 \includegraphics[width=1\linewidth]{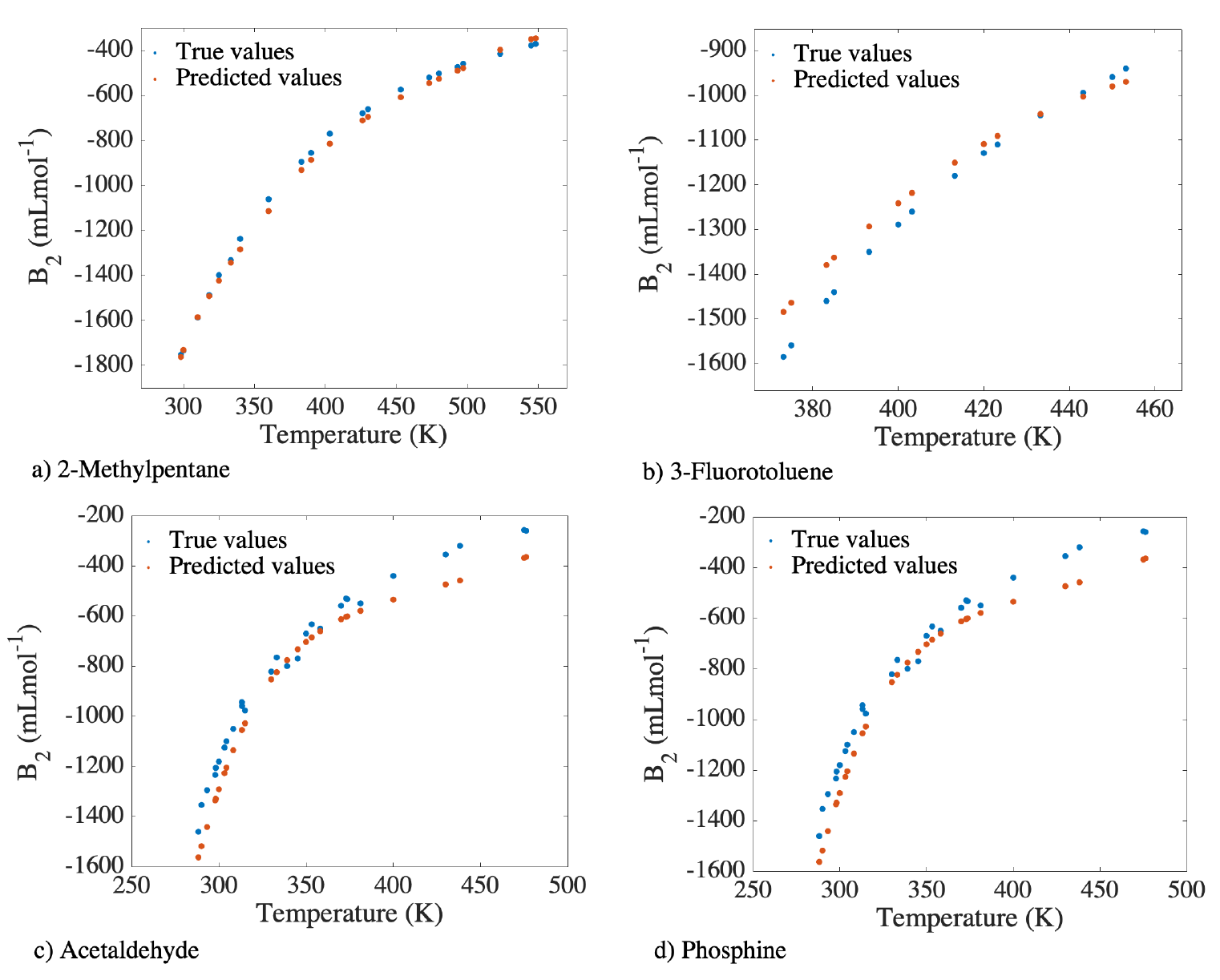}
\caption{Examples of plots showing predicted and true values for second virial coefficients as a function of temperature. Training was done on all but the presented molecules, using a GPR model with 5-fold cross validation and a 5 dimensional representation of input features. }
\label{fig:examples_predictions}
\end{center}
\end{figure}

Finally, to estimate the applicability of our model, 42 different organic molecules were left out of the training process and were tested on. This generated a training set comprising 6258 data points ($\sim 90\%$) and a test set which registered 675 data points ($\sim 10 \%$) for organic molecules only. The molecules in the test set were selected so that they cover the widest range of organic families possible (see Fig.~\ref{fig:all_data}), having, at the same time, corresponding examples of their families in the training set. The choice to only include organic molecules in the test set is rooted in the general poorer performance of inorganic compounds when compared to organic ones. While the model succeeds in predicting $B_2$ values for inorganics in interpolative regimes and when extrapolating to marginal temperatures, it is not ideally applicable to inorganic molecules which are unseen in the training set. 

Nonetheless, the model succeeds to predict second virial coefficients for organic molecules with an RMSE of 194.83 mLmol$^{-1}$ (see Fig.~\ref{fig:transferability}), which is indicative of the greater capability of the input features to describe organic compounds, rather than inorganic ones. This is naturally expected, as Morgan fingerprints incorporate valuable information for organic molecules concerning topology, element types and atomic charges. Fig.~\ref{fig:examples_predictions} reveals some examples of predicted, as well as true values for $B_2$ plotted against temperature. Instances from three different organic families are presented, for which the predicted curve is smooth. At the same time, it can be seen that the inorganic molecule phosphine performs poorly compared to the other instances. 

The model was trained using 5-fold cross-validation, having a training RMSE of 61.54 mLmol$^{-1}$ and a test RMSE of 194.83 mLmol$^{-1}$ (relative error 2.88\%). The inset of Fig.~\ref{fig:transferability} shows the distribution of the residuals from both processes and is an indication of the limitations of the model. Generally, the model predicts with great accuracy second virial coefficients for molecules well-represented in the training set, such as hydrocarbons (see Fig. 7a). However, the model is not transferable to molecules which have no resemblance to the training set, being limited to "out-of-sample" compounds that interpolate. Nonetheless, the wide range of families present in our database offer an optimistic view towards the applicability of our model, as various commonly encountered classes of organic compounds are encompassed.

\section{Conclusions}

We have developed a method for estimating second virial coefficients using Gaussian process regression with a relative error $\lesssim 1\% $ in the interpolative regime. The same model was used to predict second virial coefficients for marginal temperatures of all compounds (relative error 2.14\%), as well as for "out-of-sample" organic molecules resembling the training set (relative error 2.66\%). This has been possible through the use of a low-dimensional representation of predictors based on accessible, intuitive, and reproducible molecular features, conveniently obtained through RDKit. The applicability of our model is characterized by great performance for molecules well-represented in the training set by instances of their families, which are high in variety. When compared to traditional techniques used to calculate second virial coefficients, our method stands out in particular through its simplicity and through its efficiency, avoiding the difficulties posed by computational cost or by experimental obstacles. The input features are readily obtained through RDKit and the time required to train our best model is approximately 74 seconds on a 2 GHz Intel Quad-Core i5 machine. Finally, it is worth emphasising the important role  the existence of a comprehensive and high quality database has played in this work.

\section*{Conflicts of interest}
There are no conflicts to declare.

\section*{Appendix. Details on the GPR model}

In this work, the Gaussian process we used was defined by a rational quadratic kernel function:

\begin{equation}
\label{eq6}
K(\textbf{x},\textbf{x}')=\sigma^2\left[1+\frac{(\textbf{x}-\textbf{x}')^2}{2\alpha l^2}\right]^{-\alpha},
\end{equation}
where $\sigma^2$ is the signal variance, $l$ is the characteristic length scale of the function and $\alpha$ determines the weighting between different length scales. $l$, $\alpha$ > 0.

The ARD rational quadratic kernel function was also used:
\begin{equation}
\label{eq7}
K(\textbf{x},\textbf{x}'|\theta)=\sigma^2\left[1+\frac{1}{2\alpha}\sum_{m=1}^{d}\frac{(\textbf{x}_m-\textbf{x}'_m)^2}{\sigma^2_m}\right]^{-\alpha},
\end{equation}
where $\theta_m=\log(\sigma_m)$ for $m=1, 2,...d$ and $\theta_{d+1}=\log(\sigma)$. $d$ is the total number of predictors.

\section*{Acknowledgements}
We thank Prof. Gerard Meijer and Dr. Stefan Truppe for reading the manuscript and for useful suggestions to improve it, as well as Xiangyue Liu for fruitful discussions and for great recommendations regarding the error estimation.

\bibliography{aipsamp}

\end{document}